\documentclass[sigconf,natbib=true,anonymous=false]{acmart}

\usepackage{multirow}
\usepackage{graphicx}
\usepackage{{booktabs}}
\usepackage{tabularx}
\usepackage{textgreek}
\usepackage{subcaption}
\usepackage[notransparent]{svg}

\renewcommand\footnotetextcopyrightpermission[1]{}

\setcopyright{none}

\newcommand{\revise}[1]{\textcolor{blue}{#1}}

\AtBeginDocument{%
  }

\begin{document}

\title[]{Search or Chat? Comparing How We Learn About Debated Topics }

\author{Ran Yu}
\email{ran.yu@gesis.org}
\affiliation{%
  \institution{GESIS -- Leibniz Institute for the Social Sciences}
  \city{Cologne}
  \country{Germany}
}

\author{Alisa Rieger}
\email{alisa.rieger@gesis.org}
\affiliation{%
  \institution{GESIS -- Leibniz Institute for the Social Sciences}
  \city{Cologne}
  \country{Germany}
}

\author{Rabia Karatoprak Ersen}
\email{Rabia.KaratoprakErsen@gesis.org}
\affiliation{%
  \institution{GESIS -- Leibniz Institute for the Social Sciences}
  \city{Cologne}
  \country{Germany}
}

\author{Jiqun Liu}
\affiliation{%
  \institution{University of Wisconsin--Milwaukee}
  \city{Milwaukee}
  \state{Wisconsin}
  \country{USA}
}


\newcommand{\nb}[3]{
  \fcolorbox{black}{#2}{\bfseries\sffamily\scriptsize#1}
    {\sf\small$\blacktriangleright$\textit{\textcolor{blue}{#3}}$\blacktriangleleft$}
}

\begin{abstract} 

As large language models (LLMs) become more integrated into everyday information platforms, chat-based systems are emerging as a popular alternative to traditional web searches, especially for informational search and informal learning tasks. Despite this shift, little is known about how different tools affect learning outcomes. Our work aims to improve the understanding of how chat-based information access supports and impacts learning performance in informal learning settings. In this paper, we present the results of a crowdsourcing user study ($N = 194$) that compares learning about debated topics using a traditional search interface versus an LLM-powered chat interface. Through our analysis of learning outcomes, user characteristics, and interaction patterns, we found no significant differences in user learning gain or critical reflection on our study tasks. Our observations from the analysis of further exploratory variables suggest that, in the context of longstanding debated topics, user characteristics such as their attitude strength and level of intellectual humility might be more important in shaping immediate learning outcomes than the information access tool.
\end{abstract}

\maketitle
\pagestyle{empty}

\section{Introduction}
\label{sec:intro}

In the last decades, Web search has often been used as a starting point to learn. With the increased ease of access to LLMs and their integration to many online information platforms in easy-to-use forms such as ChatGPT, engaging with information via chat has become a normal part of many people's information access behavior. A large portion of the information behavior, especially information search with learning intent, has shifted from search to chat~\cite{yu2025chat}. However, the effect and consequences of this shift have not yet been fully understood. 
It is therefore important to investigate the impact of LLM-based tools on human information and learning behavior in order to understand how their advantages can be fully leveraged while mitigating potential negative effects on individuals and society.

Previous research in search as learning (SaL), a research field that investigates how users engage in web search activities to acquire new knowledge and develop understanding \citep{urgo2025search,marchionini2006exploratory,vakkari2016searching, rieh2016towards}, has examined the relationship between user features, task characteristics, user–system interactions, and learning outcomes. 
Important insights include that users’ effort spent on browsing web pages and their ability to formulate complex search queries are associated with learning outcomes and experiences \citep{yu2018predicting}, and that users who use more technical terms in search queries tend to achieve better learning outcomes \citep{bhattacharya2019measuring}.
However, such findings cannot be directly applied to assessing or enhancing learning in chatbot conversations.
%

With the growing usage of LLM-based chatbots, research on learning in chat has also attracted attention from IR and human-computer interaction (HCI) communities. Several research works have designed studies to investigate differences in user experience when comparing search and chat, e.g., \citep{divekar2024choosing, xu2023chatgpt}. In these studies, many participants recognized the efficiency of learning with conversational AI tools; however, they also raised concerns regarding the trustworthiness of the information and hallucination risks, and noted that it may hinder active evaluation of information and thus impair abilities such as creativity. In recent work by \citet{melumad2025experimental}, the authors compared the depth of learning achieved through web search versus LLM-based tools and found that participants developed shallower knowledge when using LLMs. This effect is particularly critical for complex, non-factual topics that require deeper cognitive processes, such as analysis, evaluation, and creation, rather than simple one-query information recall. This finding motivates us to examine the impact of chat-based information access tools on learning in the context of \textbf{debated topics}, such as “social media is good for our society”, where depth of understanding and critical reasoning are especially important~\cite{liu2021deconstructing, wang2024cognitively, kelly2015development, rieger_responsible_2024}. Our focus is on learning outcomes, including learning gain and critical reflection, while attitude change is tracked as a moderating factor rather than treated as a primary outcome. 
Existing studies on differences between search engine and LLM based chatbots mainly conduct analyses based on users’ self-reported subjective measures, without quantitative analysis of learning outcomes.
To address this gap, we conducted a crowdsourcing study to investigate the following research question:
\textbf{How does the tool of information access (search engine vs. chatbot) affect learning outcomes on debated topics?} 
%

In addition to the effect of information access tool on learning outcomes, we explore the role of topics and user characteristics, participants' information interaction behavior, the impact on users' attitudes, and their evaluation of the tool and information.  
The study employs a between subject design, in which participants were assigned a topic and one of two custom-made information access tools (traditional search interface, chat interface). We asked them to inform themselves on the assigned topic with the assigned tool. Their queries, clicks, and chat interactions were logged. 
We asked participants to complete a questionnaire prior to and after the main task, asking questions to capture their demographics, topic attitude, intellectual humility, perceived learning, and their evaluation of the information engaged with and tool used during the main task. The studies and recruitments were conducted online via Prolific\footnote{Prolific: \url{https://www.prolific.com/}\label{foot:prolific}}.

Through our work in this paper, we make the following contributions to the current body of literature:
\begin{itemize}
    \item We present a preregistered user study with 194 participants to assess the effect of using chat versus search as information access tool on learning outcomes for debated topics. The learning outcome is captured by metrics designed for measuring learning gain (\textit{argument expansion}) and critical reflection capability (\textit{critical reasoning}). To the best of our knowledge, this is the first study that quantitatively evaluates learning outcomes and examines the impact of chat on users’ cognitive skills in an informal learning context. Our dataset with learning outcome annotations is available online (see Footnote~\ref{footnote:repository}). 
    \item 
   We present findings showing no significant differences in \textit{critical reasoning} or \textit{argument expansion} between users of search engines and an LLM-based chatbot. Additionally, we provide exploratory insights indicating that chatbot users spent more time engaging with the information access tool compared to search engine users, and that to some extent, learning outcomes seem to be shaped by users' attitude strength and level of intellectual humility. We highlight directions for future research to build on these insights such as investigating emerging and more complex debated topics.
    \item We offer actionable reflections and lessons learned on meaningful assessments of learning in chat-based IR systems, highlighting the critical role of effort-sensitive evaluation methods to accurately capture the nuances of user learning.

\end{itemize}


\section{Related Work} \label{sec:relatedwork}


\textbf{Learning in Web search.}
The SaL research field emerged at the intersection of information seeking and retrieval, human-computer interaction, and education sciences~\citep{rieh2016towards, urgo2025search}. It investigates how users engage in search activities to acquire new knowledge \citep{marchionini2006exploratory,vakkari2016searching}. Research areas in SaL include user modeling, search behavior analysis, learning assessment, system support for learning, and task design \citep{von2022search,urgo2025search,vakkari2016searching}.
To better understand the learning process, studies have been conducted both online and offline to collect data. The data typically involve tracking user interactions and assessing learning outcomes. Knowledge assessment methods include quizzes (e.g. \citep{gadiraju2018analyzing}), self-reports (e.g. \citep{capra2018effects,collins2016assessing,ghosh2018searching}), essay writing (e.g. \citep{otto2022sal}), and implicit measures such as query complexity \citep{chi2016exploring}. Based on user-system interactions, researchers have also attempted to build models for predicting users' learning outcome such as knowledge gain \citep{yu2018predicting,bhattacharya2019measuring,yu2021topic,otto2021predicting}. 
%
To understand the relationship between user–system interactions and learning outcomes in search, researchers have investigated interaction features such as query complexity \citep{eickhoff2014lessons,bhattacharya2019measuring}, topic familiarity \citep{gadiraju2018analyzing}, effort-related metrics such as time spent browsing web content \citep{yu2018predicting}, and multimedia content features \citep{otto2021predicting}.

However, existing studies primarily focus on user interactions and features specific to search engines. As a result, insights into the relationships between these factors and learning outcomes cannot be directly transferred to chat-based information environments \citep{yu2025chat}. Therefore, an exploratory study of interaction features in chat-based information access scenarios is needed.
The instruments used to measure learning outcomes in prior work also informed the study design and evaluation metrics in this research. We employ questionnaires to collect subjective user measures and written essays to assess topic-related knowledge. 


\textbf{Learning with LLM-based chatbots.}
Recent research investigates how LLM-based chatbots, particularly ChatGPT, are reshaping information-seeking and informal learning behaviors~\citep{lo2023impact, zhu2023large, li2025matching}. 
\citet{xu2023chatgpt} conduct an online survey with a between-subjects design comparing ChatGPT and Google Search for information seeking tasks. They find that users complete tasks more quickly with ChatGPT and perceive its responses as higher in information quality and user satisfaction. However, ChatGPT struggles with fact-checking tasks and may induce overreliance or propagate misinformation in their study. 
\citet{divekar2024choosing} explore how higher education students perceive learning through LLM-based chatbots like ChatGPT compared to traditional search engines. Using a within-subjects study design and self-reported experience surveys, they find that students appreciate ChatGPT's conversational interface for its ease of use, scaffolded information, and synthesis capabilities, but express concerns about hallucinated information, lack of source transparency, and challenges in crafting effective prompts. 
\citet{melumad2025experimental} investigated the effects of using LLMs versus traditional web searches on the depth of learning. Their results from online and lab studies (N = 10,462) confirmed their hypothesis that participants developed shallow knowledge from LLM summaries. 
\citet{karunaratne2023new} conduct survey with students in higher education to investigate the impact of ChatGPT on their information seeking behavior. They find that ChatGPT is frequently used for information retrieval due to its perceived efficiency. However, a small portion of students express concerns regarding trustworthiness, relevance, and technical access issues. 
\citet{opara2023chatgpt} explores the challenges associated with using ChatGPT in educational settings. The authors emphasize ChatGPT's advantages, including its ability to provide quick responses to queries, support research through natural language processing and text generation, and facilitate self-paced learning. However, they also note several key limitations, such as trustworthiness, inconsistent responses to similar questions, the potential to hinder creativity, and concerns over plagiarism. 
\citet{hidayat2024examining} conduct a study to examine how AI competence, chatbot usage, and perceived autonomy influence student engagement in informal learning. 
Further research on the adoption of conversational systems in formal education settings (e.g., \citet{chakraborty2024generativeaimoderneducation,Milana02012024,Gupta2022,liu2024analysis}), including scenarios such as language learning~\citep{doi:10.1177/18479790231176372} and collaborative learning~\citep{neto2019chatbot}, also offers valuable insights applicable to CaL in informal contexts. 


Although prior work provides valuable insights into LLM-based chatbots for information seeking, evaluation, and learning, several gaps remain. Most studies rely primarily on self-reported perceptions rather than quantitative assessments of learning outcomes \cite{lee2025impact}. Moreover, little is known about how chat-based information access influences higher-order cognitive skills, such as critical reflection and reasoning. Such skills are particularly important when engaging with information on debated topics~\cite{tanprasert2024debate, rieger_responsible_2024}. 
To address these gaps, our study quantitatively evaluates learning outcomes and examines the impact of chat-based interaction on users’ critical reasoning in an informal learning context.

\section{Method}
\label{sec:method}


We conducted a preregistered\footnote{The preregistraion, in which we described our hypotheses and study design prior to data collection, can be found in our repository: \url{https://osf.io/wh9j5/overview?view_only=63406cd6e1324b539b355e1e51f441fe}\label{footnote:repository}} crowdsourcing user study with a one-way between subjects design to investigate how the tool of information access affects learning outcomes on debated topics.

\subsection{Hypotheses}\label{sec:hypotheses}
To answer the question "How does the tool of information access (search engine vs. chatbot) affect learning outcomes on debated topics?", we tested the following hypotheses:

        \begin{itemize}
            \item \textbf{H1:} Participants who interact with the chatbot will show less argument expansion than participants who interact with the search engine. 
            \item \textbf{H2:} Participants who interact with the chatbot will show less critical reasoning than participants who use search engines. 
        \end{itemize}

\subsection{User Study Procedure}
\label{sec:procedure}

\begin{figure*}[!t]
  \centering

  \begin{subfigure}[t]{0.8\textwidth}
    \centering
    \includegraphics[width=\linewidth]{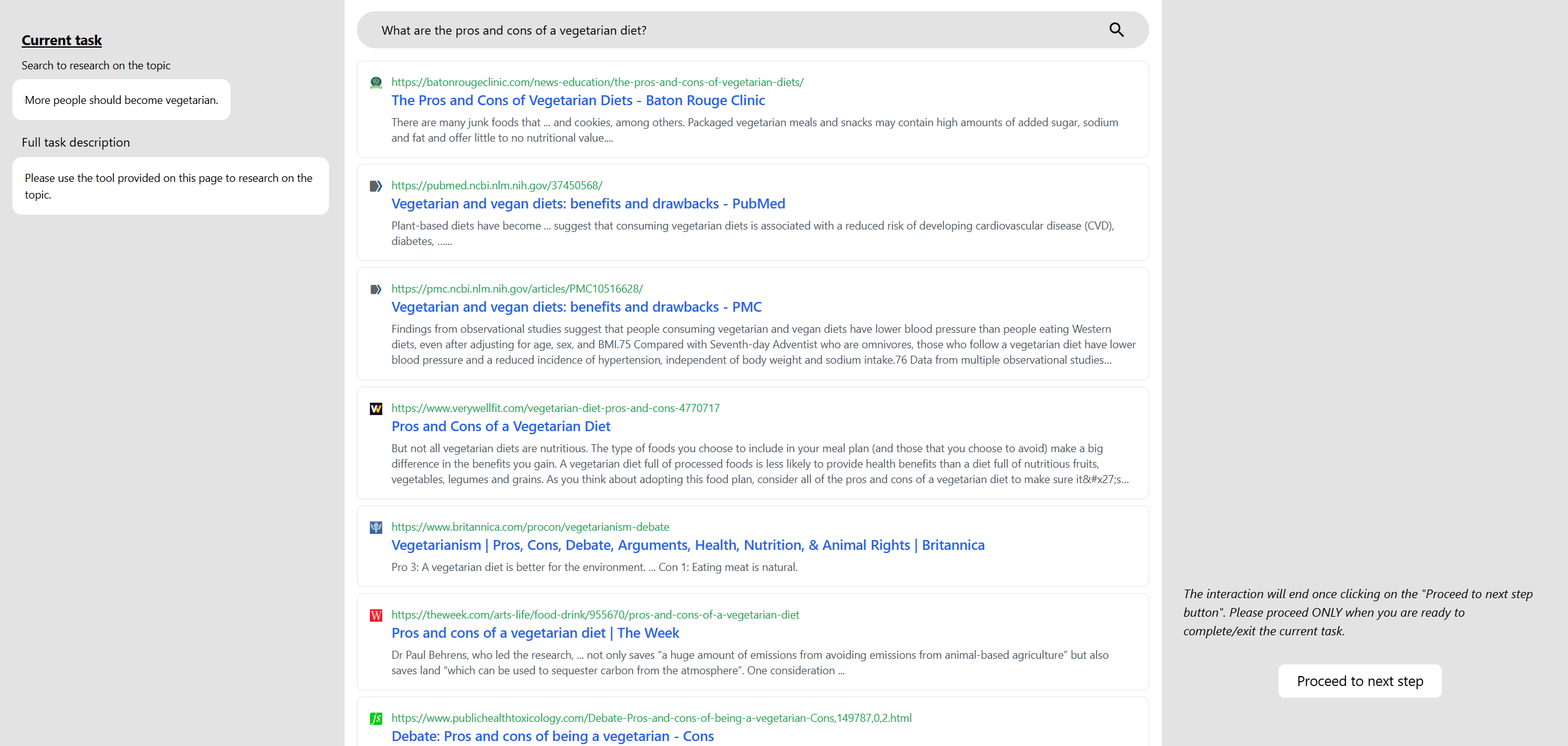}
    \caption{Search}
    \label{fig:search-page}
  \end{subfigure}

  \vspace{0.8em}

  \begin{subfigure}[t]{0.8\textwidth}
    \centering
    \includegraphics[width=\linewidth]{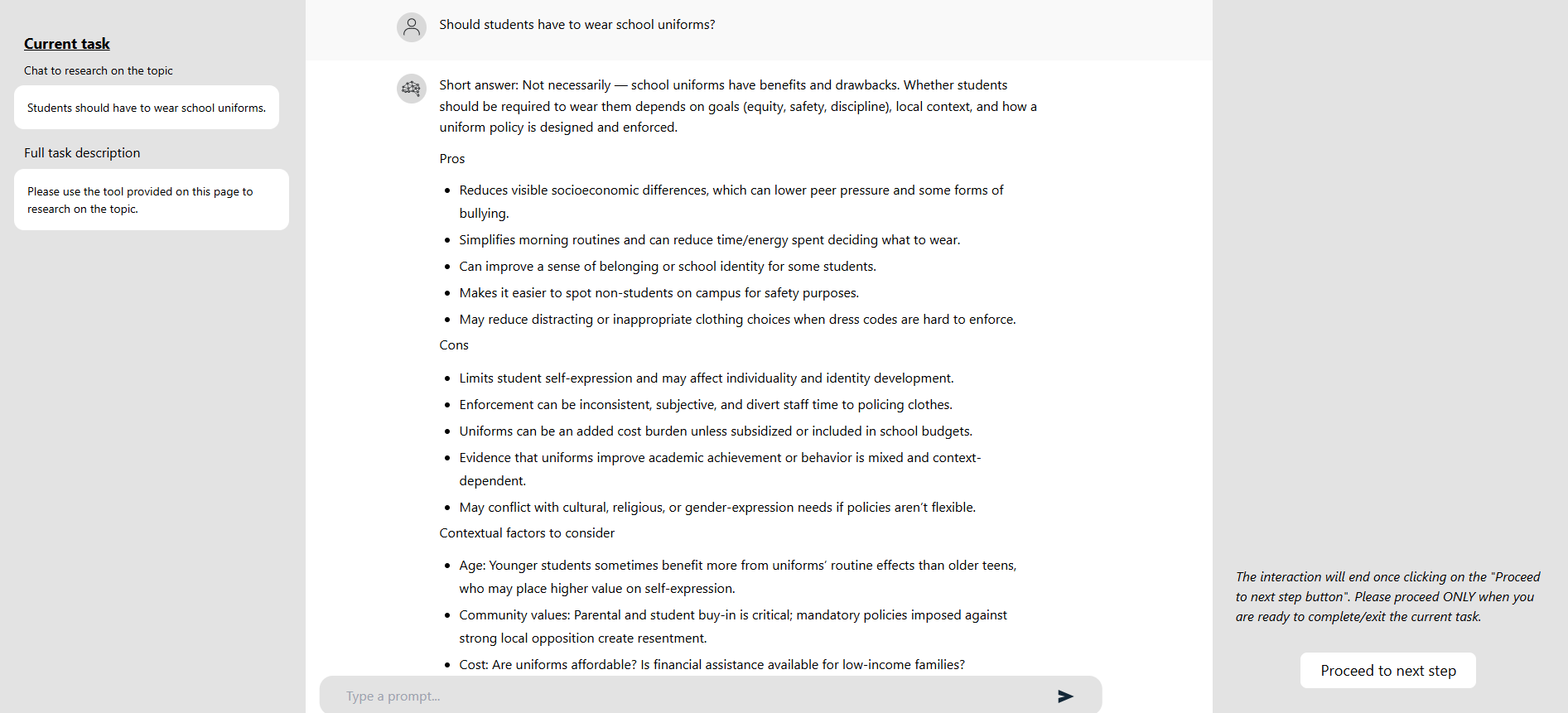}
    \caption{Chat}
    \label{fig:chat-page}
  \end{subfigure}

  \caption{The search and chat interface.}
  \label{fig:tool-page}
\end{figure*}

\begin{figure}[!t]
  \centering
  \includegraphics[
    width=\columnwidth,
    trim=80 0 80 30,
    clip
  ]{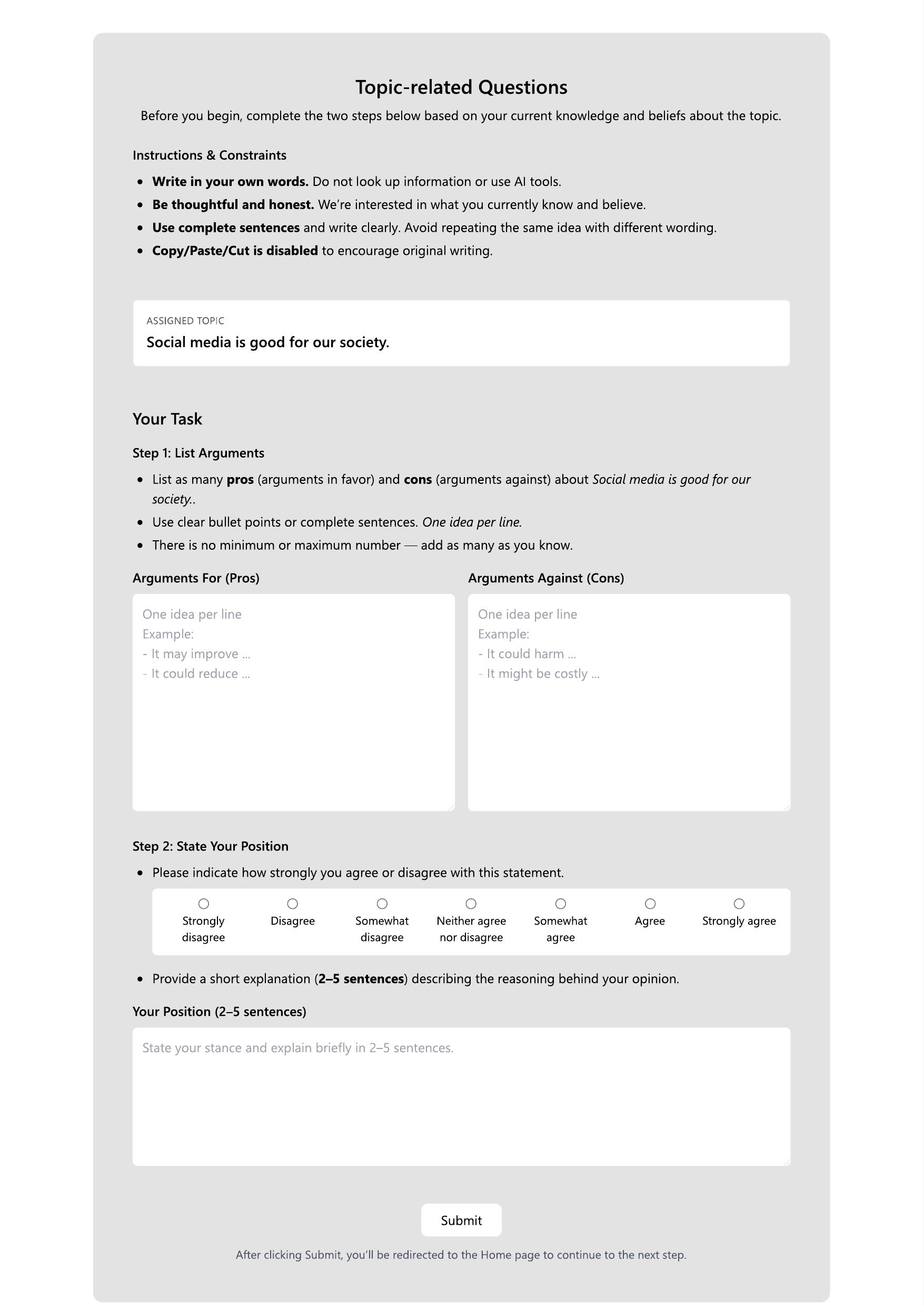}
  \caption{Topic-related writing task.}
  \label{fig:essay-page}
\end{figure}

We collected data via an online platform customized to our needs based on an open-source project\footnote{\url{https://github.com/JamshedK/fixednessEvalution/tree/full_study_update}}. The custom made search engine (see Figure \ref{fig:search-page}) is based on the Brave API\footnote{\url{https://brave.com/search/api/}}, and the chatbot (see Figure \ref{fig:chat-page}) on OpenAI\footnote{\url{https://openai.com/api/}} model gpt-5-mini. We recruited participants via Prolific\footref{foot:prolific}. They were required to be at least 18 years old, and English had to be the primary language. Participants completed the study in five steps, following the procedure described below, which has been approved by the ethics committee of our institution. When completing the study, participants were rewarded with \pounds 3.39.
\begin{itemize}
    \item \textbf{Step 1.} After providing informed consent to participate in this study, 
    we randomly assigned them to one of the following topic statements: \textit{Bottled water should be banned}, \textit{Social media is good for our society}, \textit{More people should become vegetarian}, \textit{Students should have to wear school uniforms}. They were also assigned to one of the two information access tools (search engine, chatbot).
    \item \textbf{Step 2.} For the assigned topic, we asked participants to list pro and con arguments (see Figure \ref{fig:essay-page}). Further, we asked them to report their attitude on the topic on a seven-point Likert scale, ranging from \textit{strongly disagree} to \textit{strongly agree}. Subsequently, we asked them to provide a short explanation of their position. 
    \item \textbf{Step 3.} In the next step, participants were redirected to the assigned information access tool (see Figure~\ref{fig:tool-page}). They were instructed to use the tool to research the assigned topic. Participants could have as many interactions as they wanted but had to issue at least one prompt/query before they were able to proceed to the next step, getting redirected to the survey interface. 
    \item \textbf{Step 4.} Participants again were asked to list pro and con arguments, report their attitude on the topic, and provide a short explanation for it (see Step 2).
    \item \textbf{Step 5.} In the final step, we asked participants to answer questions about their attitude certainty, their perceived learning, and their experience with the information and the tool. They were also asked to answer questions from the intellectual humility scale~\cite{leary_cognitive_2017}, and could provide feedback on the task.
\end{itemize}

 To ensure good data quality, participants had to pass three attention checks in which they were instructed which response to select, one in the pre- and two in the post-task questionnaire.

\begin{figure}[!t]
    \centering
    \includesvg[width=1\linewidth]{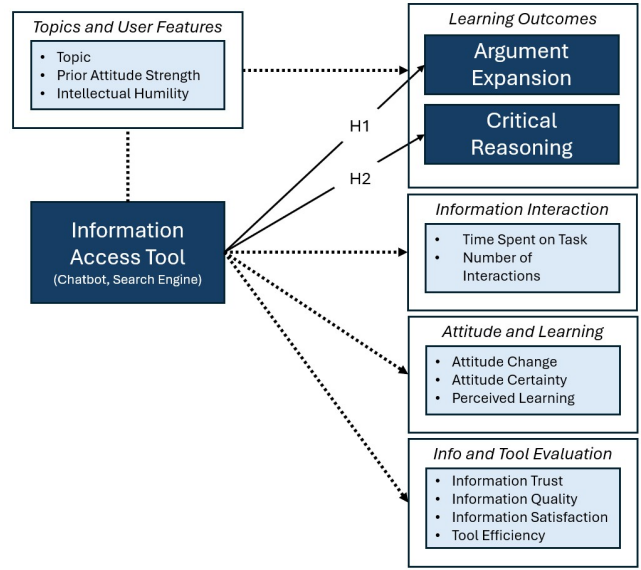}
    \caption{Overview of independent and dependent variables (dark blue) and additional exploratory variables (light blue); arrows indicate relationships tested in hypothesis testing (solid lines) and exploratory analysis (dotted lines).}
    \label{fig:variable_overview}
\end{figure}

\subsection{Variables}
\label{sec:variables}

In this section, we describe our independent variables, dependent variables for the hypotheses defined in Section \ref{sec:hypotheses} in detail. To expand our understanding of accessing information on debated topics with chatbots compared to search engines, we investigate additional exploratory variables beyond those included in our hypotheses. Figure~\ref{fig:variable_overview} provides an overview of all variables.

\paragraph{\textbf{Independent variable. }} Our primary objective is to compare the effects of search engines and chatbots on learning outcome on debated topics. Accordingly, we define the \textit{information access tool} as the independent variable.

\begin{itemize}
    \item \textbf{Information Access Tool} (between-subjects, categorical) [chatbot, search engine]. Participants were assigned to one of the two information access tools, the search engine or the chatbot. They were asked to use the assigned tool to find information on the assigned topic (see Section~\ref{sec:procedure}).
    \end{itemize}

\paragraph{\textbf{Dependent variables.}} As quantitative measures of learning outcomes about debated topics, we analyze participants’ written responses in Step 2 and Step 4 of the study (Section~\ref{sec:procedure}). To measure learning outcomes, we rely on open text responses instead of knowledge quizzes, to avoid inadvertently influencing participants' information interactions with the content of the quiz questions. Moreover, designing objective knowledge tests for contested topics poses significant methodological challenges.
Instead, we asked participants to report pro arguments, con arguments, and to provide an explanation of their positions in free text fields. From their responses, we compute two metrics: (1) \textit{argument expansion}, the number of novel arguments reported after the task, which we interpret as an approximation of participants' learning of arguments in the debate, and (2) \textit{critical reasoning}, derived from the explanations of their positions, particularly their use of arguments, based on which we can approximate their critical understanding of the broader debate. 

\begin{itemize}
    \item \textbf{Argument Expansion} (integer). Argument Expansion was calculated as the number of novel pro and con arguments reported after the information task that had not been reported before the information task\footnote{This is a deviation from the preregistration in which we planned to simply calculate the difference in the number of arguments reported before and after. However, we observed in that most participants did not repeat the arguments they reported before the task, so the preregistered difference score was not meaningful. We therefore adapted the metric to capture \textit{argument expansion}.}. To calculate argument expansion, three investigators independently coded an initial set of 20 responses and met to establish consensus on criteria that indicate distinct and novel arguments. The criteria were provided to GPT-5 mini, which we used as a rule-based classifier\footnote{The prompts and coding schemes for the \textit{argument expansion} and \textit{critical reasoning} scores can be found in our repository, linked in Footnote~\ref{footnote:repository}\label{foot:prompts}}, prompted for batches of 10 participant to provide the counts. GPT-generated counts were reviewed by one of the investigators who routinely verified outputs to ensure that the arguments identified matched the arguments provided. To assess GPT-rater reliability, one of the investigators manually recoded a random sample of 20 responses. We observed high agreement between GPT and the coder with Krippendorff's $\alpha = 0.91$ for the argument expansion score.
    
    \item \textbf{Critical Reasoning.} (ordinal) [0-4]. 
    The level of \textit{critical reasoning} was determined by evaluating participants' responses from Step 4 of the study when instructed to \textit{provide a short explanation (2–5 sentences), describing the reasoning behind their opinion}.
    The free-text responses were evaluated with a coding scheme, based on the \textit{reasoning using evidence, argumentation, and synthesis} facet of~\citet{molerov_assessing_2020}'s assessment framework. The coding scheme we applied consisted of the features \textit{coherence} (0 to 2 points) and \textit{perspectives mentioned} (0 to 2 points), summed as the overall \textit{critical reasoning} score. 
    Participants, who have a critical understanding of the broader debate, should be able to base their arguments on plausible reasons. Coherence together with mentioned perspectives represent their reasoning skill as supported by ~\citet{molerov_assessing_2020}'s assessment framework. 
    Three investigators independently coded an initial set of 20 responses and then met to establish consensus, refining the coding scheme as needed. This final scheme and 5 examples from the manually coded responses were provided to GPT-5 mini to annotate the remaining dataset\footref{foot:prompts}. 
    As for argument expansion, one of the investigators manually recoded a random sample of 20 responses, and found perfect agreement of Krippendorff's $\alpha = 1$ between GPT and the coder for the coherence and the perspectives scores. 
\end{itemize}


\paragraph{\textbf{Exploratory variables.}} We group our exploratory variables according to their role in the learning process into four categories:
1) \textit{topics and user factors}, which serve as static inputs to the learning process;
2) \textit{information interaction}, which is captured during participants’ interactions with the search engine or chatbot;
3) \textit{attitude and learning}, which are coded from participants’ written assignments; and
4) \textit{information and tool evaluation}, which are self-reported measures collected after the learning process. An overview of the exploratory variables can be seen in Figure \ref{fig:variable_overview} (light blue boxes). 
Below, we describe their definitions and calculation procedures.
\begin{itemize}
    \item \textbf{Topic}. One of \textit{Bottled water should be banned}, \textit{Social media is good for our society}, \textit{More people should become vegetarian}, \textit{Students should have to wear school uniforms}, randomly assigned between subjects.
    \item \textbf{Prior Attitude Strength} [0-3]. Questionnaire response in Step 2 to \textit{Please indicate how strongly you agree or disagree with this statement} on a 7-point Likert scale from \textit{strongly disagree} to \textit{strongly agree}, mapped to a 4-point scale from \textit{undecided} to \textit{strong attitude}.
    \item \textbf{Intellectual Humility} [1-5]. Intellectual humility refers to the extent to which people recognize the fallibility of their own beliefs and remain open to revising those beliefs in light of new evidence or alternative viewpoints~\cite{porter_predictors_2022}, a user trait we measured with the \textit{intellectual humility scale} by~\cite{leary_cognitive_2017} (see Section~\ref{sec:procedure}, Step 5). For exploration of group differences, scores were categorized based on the quartiles of the distribution across all participants. Values within Q1 were classified as \textit{low}, those within Q2 and Q3 as \textit{moderate}, and those within Q4 as \textit{high}.
    \item \textbf{Time Spent on Task} (integer). Time spent interacting with the information access tool measured in seconds, as the time difference between participants first accessing the tool and proceeding to the post-task questionnaire. 
    \item \textbf{Number of Interactions} (integer). Number of prompts in the chatbot condition, and number of queries and clicks, which we considered separately and combined, in the search engine condition. 
    \item \textbf{Attitude Change} [unchanged, strengthened, weakened, flipped]. Difference between attitude on the topic reported before and after engaging with the information access tool on a 7-point Likert scale, mapped to the outcomes\textit{unchanged, strengthened, weakened} or \textit{flipped}. 
    \item \textbf{Attitude Certainty} [1-5]. Questionnaire response to \textit{How certain are you about your opinion on the topic you just engaged with?} on a 5-point Likert scale from \textit{very uncertain} to \textit{very certain}.
    \item \textbf{Perceived Learning.} [1-5]. Questionnaire response to \textit{To what extent do you feel that you learned something new or increased your understanding from using this tool?} on a 5-point Likert scale from \textit{not at all} to \textit{very much}.
    \item \textbf{Information Trust} [1-5]. Questionnaire response to \textit{How much do you trust the information you received from this tool?} on a 5-point Likert scale from \textit{not at all} to \textit{completely}.
    \item \textbf{Information Quality} [1-5]. Questionnaire response to \textit{How would you rate the overall quality of the information you received?} on a 5-point Likert scale from \textit{very poor} to \textit{excellent}.
    \item \textbf{Information Satisfaction} [1-5]. Questionnaire response to \textit{Overall, how satisfied are you with your experience using this tool for the assigned task?} on a 5-point Likert scale from \textit{very dissatisfied} to \textit{very satisfied}.
    \item \textbf{Tool Efficiency} [1-5]. Questionnaire response to \textit{How efficient was the process of getting the information?} on a 5-point Likert scale from \textit{very inefficient} to \textit{very efficient}.
\end{itemize}

\section{Results and Analysis}
\label{sec:results}
Below, we report the results from testing our preregistered hypotheses, and additional insights gained from data explorations beyond our hypotheses. 

\subsection{Description of the Sample}
With an apriori power analysis, we determined a required sample size of 202 participants, expecting a moderate effect ($f = 0.25$), a significance threshold \textalpha~= $\frac{0.05}{2}$ = 0.025 due to testing two hypotheses, a desired power of (1-~\textbeta)~= 0.9 and planning to test two groups (i.e., search engine, chatbot) with between-subjects ANOVAs. 
Initially, 219 participants completed the study, of which 23 failed one or more attention checks, and two submitted invalid questionnaires (i.e., reporting off-topic arguments). Consequently, 194 participants were included in the data analysis.
Of these, 116 reported to be female, 75 male, and 3 non-binary/other. Regarding their age, 29\% reported to be between 18 and 25, 33\% between 26 and 35, 20\% between 36 and 45, 15\% between 46 and 65, and 3\% above 65 years old. 
70\% reported to be residing in Sub-Saharan Africa (68\% in South Africa), 17\% in Europe and Central Asia, 6\% in North America, and the remaining 7\% across multiple other regions. 
The participants were mostly highly educated, with 12\% reporting to hold a PhD, 21\% a Master's degree, 54\% a Bachelor's degree, 6\% other college degrees, and 7\% a Highschool degree. 

\subsection{Hypothesis Testing}
First, we report the results of testing the hypotheses defined in Section~\ref{sec:hypotheses}.

\begin{figure*}[!t]
    \centering
    \includesvg[width=0.92\linewidth]{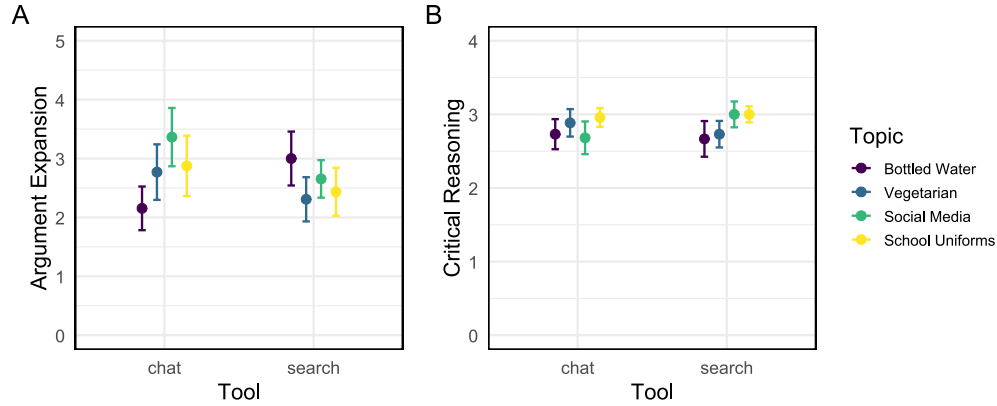}
    \vspace{-0.3 in}
    \caption{\textit{Argument Expansion} (A) and \textit{Critical Reasoning} (B) per Tool and Topic: Mean values with Standard Errors.}
    \label{fig:AE_CR}
\end{figure*}

\textbf{H1: Effect of Tool on Argument Expansion.}
Participants who engaged with the search engine reported $M = 2.58$ ($SE = 0.19$) novel arguments, and participants who engaged with the chatbot reported $M = 2.77$ ($SE = 0.23$) novel arguments. 
%
Results of an ANOVA indicate no difference in \textit{argument expansion} between search engine and chatbot ($F (1, 192) = 0.37, p = .55, f = .04$).


\textbf{H2: Effect of Tool on Critical Reasoning.}
The mean \textit{critical reasoning} level observed in the textual justifications of participants who engaged with the search engine was $M = 2.85$ ($SE = 0.09$), and of participants who engaged with the chatbot $M = 2.82$ ($SE = 0.09$).
An ANOVA indicated no difference in \textit{critical reasoning} between the search and the chatbot ($F (1, 192) = 0.085, p = .77, f = .02$).

\subsection{Exploratory Analysis}
The exploratory variables allow us to further explore potential differences between chatbot and search engine users, and to investigate factors that may moderate their learning outcomes. 
For all exploratory variables discussed in this section, we examined whether each variable moderated the effect of the information access tool on the dependent variables. To this end, we fitted regression models with the information access tool, the exploratory variable, and their interaction term as predictors. We did not observe significant interaction effects for any of these variables. In the rest of this section, we primarily report descriptive analyses on the relations indicated as dotted lines in Figure \ref{fig:variable_overview}. Insights gained from this should not be interpreted as evidence for effects, but a means to improve the understanding of the data we collected and highlight potential directions for future research.


\subsubsection{Beyond Learning Outcomes}
 To gain insights into differences between participants who engaged with the chatbot and with the search engine beyond the learning outcomes, we explored participants' information interaction behavior, attitude and learning, and information and tool experience. 

\begin{figure*}[!t]
    \centering
    \includesvg[width=0.97\linewidth]{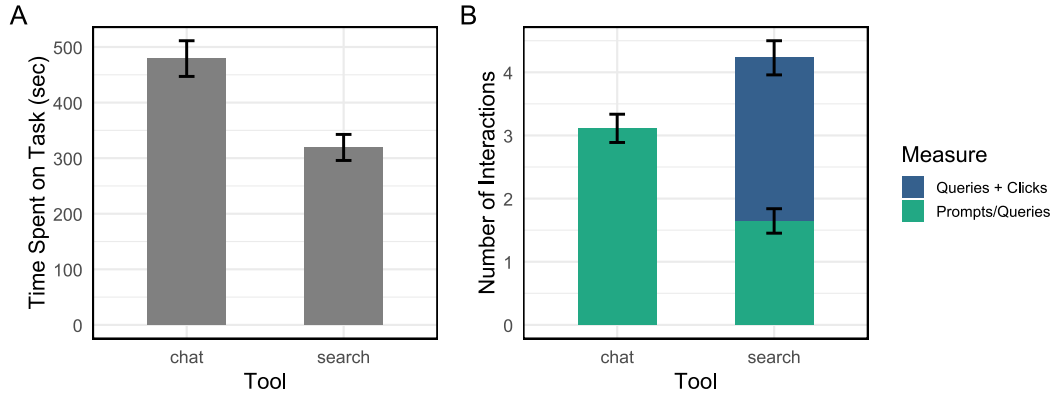}
    \vspace{-0.1 in}
    \caption{Time Spent on Task (A) and Number of Interactions (B) per Tool and Topic: Mean values with Standard Errors. Figure B shows to the number of queries (green) and of queries and clicks combined (blue) in the search engine condition.}
    \label{fig:interactions}
\end{figure*}

\textbf{Information Interaction Behavior.}
We observed that participants who interacted with the chatbot spent a higher mean time on the information seeking task ($M = 479s$, $SE = 32.1$) than participants who interacted with the search engine ($M = 329s$, $SE = 23.4$) (see Figure~\ref{fig:interactions}). 
In terms of the number of interactions, since they are not directly comparable across conditions, we provide this information only as a reference. We observed that participants in the chatbot condition had $M = 3.11$, $SE = 0.23$ prompts on average. In the search engine condition, participants had $M = 4.23$, $SE = 0.38$ interactions on average (see Figure~\ref{fig:interactions}).

\textbf{Attitude and Learning.}
In exploring attitude change, we observed similar proportions of participants who did not change, strengthened, weakened, or flipped their attitudes across the chatbot and search engine conditions (see Table~\ref{tab:att_change_cat}).
We also did not observe any differences in attitude certainty reported by participants in the chabot ($M = 4.37$, $SE = 0.07$) and in the search engine ($M = 4.36$, $SE = 0.06$) condition.
There were no differences in perceived learning in chatbot ($M = 3.85$, $SE = 0.11 $) and search engine users ($M = 3.88$, $SE = 0.11$).

\begin{table}[t]
\centering
\caption{Distribution of attitude change outcomes of participants using either the chatbot or the search engine.}
\label{tab:att_change_cat}
\begin{tabular} {llc}
\toprule
\textbf{IA tool} & \textbf{Outcome} & \textbf{Participants (\%)} \\
\midrule
Chatbot & unchanged &  66.3\% \\
 & strengthened & 16.3\% \\
 & weakened & 12.2 \% \\
 & flipped & 5.1\% \\
 \midrule
 Search Engine & unchanged &  67.7\% \\
 & strengthened & 17.7\% \\
 & weakened & 11.5 \% \\
 & flipped & 3.1\% \\
\bottomrule
\end{tabular}
\end{table}

\textbf{Information and Tool Evaluation.}
We did not observe any differences between chatbot and search engine users in the reported levels of information trust, information quality, information satisfaction, and tool efficiency (see Table~\ref{tab:info_tool_eval}).

\begin{table}[t]
\centering
\caption{Information and tool evaluation for chatbot and search engine: mean and standard error.}
\label{tab:info_tool_eval}
\begin{tabular} {lll}
\toprule
\textbf{IA tool} & \textbf{Variable} & \textbf{Mean and SE} \\
\midrule
Chatbot & Information Trust &  $M = 4.18$, $SE = 0.08$ \\
 & Information Quality &  $M = 4.37$, $SE = 0.06$\\
 & Information Satisfaction &  $M = 4.33$, $SE = 0.07$\\
 & Tool Efficiency & $M = 4.22$, $SE = 0.08$\\
 \midrule
 Search Engine & Information Trust &  $M = 4.1$, $SE = 0.08$ \\
 & Information Quality &  $M = 4.21$, $SE =0.09$\\
 & Information Satisfaction & $M = 4.38$, $SE = 0.07$\\
 & Tool Efficiency & $M = 4.16$, $SE = 0.08$ \\
\bottomrule
\end{tabular}
\end{table}

\subsubsection{Moderation by Topics and User Factors}
To gain insights in potential moderating factors of the relation between information access tool and learning outcomes, we explored differences between topics and selected user factors.

\textbf{Topic.}
While we observed minor variations of mean \textit{argument expansion} and \textit{critical reasoning} scores between topics (see Figure~\ref{fig:AE_CR}), exploratory two-way ANOVAs did not reveal a moderating effect of Topic on the relationship between information access tool and learning outcomes.

\begin{figure*}
    \centering
    \includesvg[width=2.3\columnwidth]{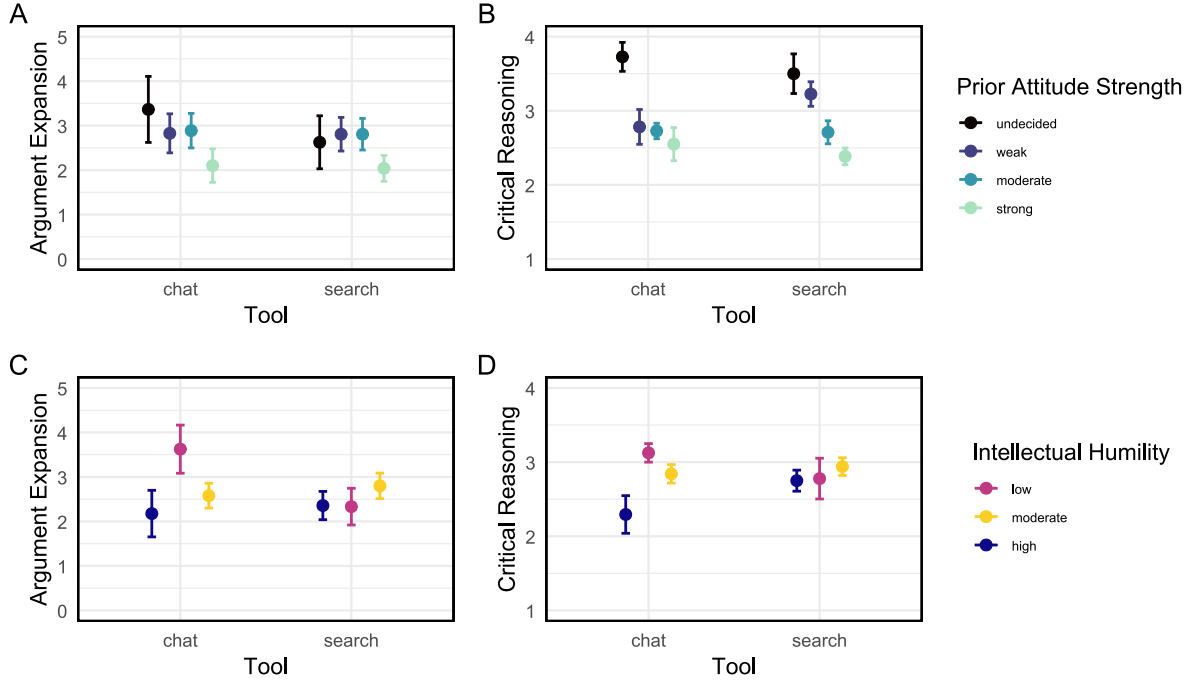}
    \vspace{-0.3 in}
    \caption{\textit{Argument Expansion} and \textit{Critical Reasoning} per Tool and Level of Attitude Strength (A, B), or Level of Intellectual Humility (C, D): Mean values with Standard Errors}
    \label{fig:attitude_IH}
\end{figure*}

\textbf{Attitude Strength.}
Attitude strength did not moderate the relationship between information access tool and learning outcomes, however, descriptive statistic shows that independent of the information access tool, participants with strong prior attitudes showed lower \textit{critical reasoning} and \textit{argument expansion} than participants with weaker attitudes, or who were undecided (see A and B in Figure~\ref{fig:attitude_IH}).

\textbf{Intellectual Humility.}
We observed a mean intellectual humility score of $M = 3.83$ ($SE = 0.06$). For participants who interacted with the search engine, we observed no differences in learning outcomes between different levels of intellectual humility. However, for participants who interacted with the chatbot, our data indicates a difference in learning outcomes between different levels of intellectual humility, where participants with low intellectual humility showed higher \textit{argument expansion} and \textit{critical reasoning} than participants with high intellectual humility (see Fig.~\ref{fig:attitude_IH}).

\section{Discussion}
\label{ref:discussion}
With this user study ($N = 194$), we investigate differences between chatbot and search engine users who engage with information on debated topics in their learning outcomes. We also explored  differences in their information interaction behavior, attitudes and tool and information evaluation. Additionally, we explored the role of topics, attitude strength, and intellectual humility in shaping learning outcomes of chatbot and search engine users.

\subsection{Findings and Implications}

\textbf{Tool Differences in Learning Outcomes.}
We observed no differences between chatbot and search users in the number of novel arguments they reported after engaging with information, or the level of \textit{critical reasoning} they showed when justifying their attitudes. Consequently, we did not find evidence to support either \textbf{H1}, an effect of information access tool on \textit{argument expansion}, or \textbf{H2}, an effect of information access tool on \textit{critical reasoning}.


\textbf{Tool Differences Beyond Learning Outcomes.}
Exploring differences in information interaction behavior, we observed that chatbot users spent more time engaging with the information access tool than search engine users. This could indicate that information interactions were more engaging with the chatbot. However, given that we did not observe better learning outcomes for chatbot users than search engine users, it is also possible that interactions with the chatbot were less efficient, for instance, because participants had to read through lengthy responses to their prompts. 
The underlying causes of this time difference should be investigated in future research.
Comparing participants' interactions revealed that the mean number of queries of search engine users was lower than the mean number of prompts of chatbot users. Considering both queries and clicks, we observed that search engine users made more active choices to complete the information task, as to be expected by the tool differences.

Our data revealed no differences between chatbot and search users when exploring participants' attitude change, attitude certainty, and perceived learning.
We also did not observe any differences in their overall positive evaluations of the tool they used to access information and the information the tool provided.

\textbf{The Role of Topics and User Factors.}
Exploring differences in participants' learning outcomes between topics, we observed some variations in means, but no systematic differences. 
Our investigation of participants prior attitude strength indicated a link to learning outcomes, in particular to \textit{critical reasoning}, where those with strong prior attitudes demonstrated lower levels of \textit{critical reasoning} than those with weak attitudes. This observation is consistent with prior findings on the role of users' attitude strength in shaping engagement with information on debated topics~\cite{rieger_disentangling_2024}. 
Further, our explorations revealed variations in learning outcomes linked to levels of intellectual humility of chatbot users, but not of search engine users. Here, chatbot users with high levels of intellectual humility showed less \textit{argument expansion} and \textit{critical reasoning} than chatbot users with low intellectual humility. These curious variations among chatbot users should be further investigated. Such research would contribute to the growing body of work on how intellectual humility shapes information interactions across information access tools~\cite{gorichanaz_relating_2022, porter_predictors_2022, rieger_potential_2024}.

\textbf{Implications.} 
Overall, our findings and observations suggest that in the context of immediate learning outcomes for longstanding debated topics, what users bring to the task, such as their attitude strength and level of intellectual humility, might be more important in shaping learning outcomes than the information access tool. 
This limited role of information access tool in shaping learning outcomes has been observed similarly in other studies, which found efficient, immediate learning and understanding among users who access information via chatbots~\cite{yang_can_2024, yang_search+chat_2025, mayerhofer_blending_2025}.
Yet, in light of recent findings from an extensive study by~\citet{melumad2025experimental}, showing that advice given by chatbot users was shallower, less invested, and less likely to be adopted by others, we want to emphasize the need for studies investigating differences beyond immediate learning, deeper topic understanding, and downstream behavior.

\subsection{Reflections on Chat as Learning Evaluation}
Evaluating learning in chat-based IR systems requires moving beyond topical relevance and self-reported perceptions, using process- and outcome-focused measures that reveal how users navigate, understand, and integrate information, and then use it to form and express their thoughts and opinions.
Building on SaL research, we operationalized learning through structured pre- and post-task writing and quantified both \textit{argument expansion} and \textit{critical reasoning}, enabling an approximation of learning of new arguments and understanding of the broader debate, rather than mere user perceptions. This design highlights an important challenge for learning evaluation in chat: demonstrating reasoning depends not only on access to information, but also on users’ motivation, effort, and willingness to externalize their thinking. Our findings show no significant differences between chat and search on either \textit{argument expansion} or \textit{critical reasoning} for longstanding debated topics, suggesting that chat-based interfaces do not inherently undermine immediate learning when evaluated with task-aligned measures. At the same time, exploratory results reveal that engagement patterns and user characteristics, such as attitude strength and intellectual humility, shape what users demonstrate as learning, particularly in chat settings. These results jointly support the value of carefully designed, effort-sensitive evaluation for studying learning in conversational IR, and caution against attributing learning effects solely to the tool without considering how reasoning is measured.



\subsection{Limitations and Future Work}

To ensure comparability between the two information access tools, we conducted a controlled user study, coming with limitations on the generalizability of our findings. For instance, we had to restrict the number and types of topics we investigated. 
We selected accessible, widely-known topics, enhancing suitability for participants from diverse backgrounds. 
Aiming for high ecological validity, we did not restrict the information that participants could access via both tools.  
By exploring the logged interactions, we observed that both the search and chat systems provided participants with generally comprehensive information that sufficiently covered the corresponding task topics. This helps ensure a fair comparison between conditions, as the primary difference between them is the interface and interaction mode of the two tools. 
However, longer-standing debated topics possibly limiting measurable learning gains due to prior topic familiarity. 
More current debates bring higher uncertainty, possibly linked to increased variations in information quality between chatbots and search engines. 
Future studies could explore less established, more complex debates, to build on and further validate our findings. 


We also acknowledge the limited geographical diversity of participants who completed our study, which limits the generalizability of our findings to users from different regions. Based on prior studies conducted via Prolific, we were surprised by this imbalance, prompting us to design for and monitor participant diversity more carefully in future studies. 

Lastly, experienced crowdsourcing workers may be sensitive to researchers’ expectations, also known as demand characteristics, which could influence their behavior and reduce observable differences between systems \citep{mccambridge2012effects}. Carefully designed field studies may therefore complement our findings.

\section{Conclusions}
\label{ref:conclusion}

We conducted a between-subjects crowdsourcing study to examine how different information access tools influence online, informal learning about debated topics. To complement primarily qualitative prior research on differences between search-based and chat-based information access for learning, we developed quantitative measures to capture learning outcomes, like participants' argument expansion and critical reflection. 
Our results indicate no differences in these learning outcomes between participants using a search engine and those using an LLM-based chat system. Exploratory analysis did, however, indicate differences between participants with different levels of attitude strength and intellectual humility, and differences between chatbot and search engine users in the time they spent engaging with the information access tool. 
These insights imply that, for longstanding debated topics and immediate learning outcomes, user factors may have a greater impact on learning than the information access tool used. More studies are needed to fully understand the impact of the ongoing shift from search engines to LLM-based chatbots for information seeking on individuals and society.




\balance
\bibliographystyle{ACM-Reference-Format}
\bibliography{9_references.bib}

\end{document}